\documentclass[aps,pra,reprint,groupedaddress,floatfix]{revtex4-1}

\usepackage{amsmath,amssymb,bm,graphicx,url}
\usepackage{bbold}
\usepackage[caption=false]{subfig}
\usepackage{color}
\usepackage{cancel}

\newcommand{\be}{\begin{equation}}
\newcommand{\ee}{\end{equation}}

\newcommand{\sz}{\sigma_z}
\newcommand{\sm}{\sigma_-}
\renewcommand{\sp}{\sigma_+}
\newcommand{\expec}[1]{\left\langle #1 \right\rangle}
\newcommand{\bra}[1]{\left\langle \, #1 \right|}
\newcommand{\ket}[1]{\left| #1  \right\rangle}
\newcommand{\lind}[1]{\mathcal{D}\left[#1\right]}

\newcommand{\comm}[2]{\left[ #1, #2 \right]}
\newcommand{\Tr}{\mbox{Tr}}

\begin{document}


\title{Simple, robust and on-demand generation of single and correlated photons}


\author{Sankar Raman Sathyamoorthy}
\email[]{sankarr@chalmers.se}
\author{Andreas Bengtsson}
\author{Steven Bens}
\author{Micha\"{e}l Simoen}
\author{Per Delsing}
\author{G\"{o}ran Johansson}
\affiliation{Department of Microtechnology and Nanoscience, MC2, Chalmers University of Technology, Gothenburg, SE-41296.}

\date{\today}

\begin{abstract}
We propose two different setups to generate single photons on demand using an atom in front of a mirror, along with either a beam-splitter or a tunable coupling. We show that photon generation efficiency $\sim 99\%$ is straightforward to achieve. The proposed schemes are simple and easily tunable in frequency.  The operation is relatively insensitive to dephasing and can be easily extended to generate correlated pairs of photons. They can also in principle be used to generate any photonic qubit of the form $\mu \ket{0} + \nu\ket{1}$ in arbitrary wave-packets, making them very attractive for quantum communication applications.
\end{abstract}


\maketitle


\section{Introduction}
Single photons, the fundamental excitations of the electromagnetic field, are prime candidates to transfer quantum information across a network. As opposed to coherent fields (even at weak powers), single photons and the higher number states are clearly non-classical as can be seen in interference\cite{Grangier1986} and correlation measurements\cite{Hong1987}. The non-classical nature of single photons can not only be used to test the foundations of quantum mechanics\cite{Aspect1981} but also for several applications in quantum communication\cite{Gisin2002}, quantum computing\cite{Knill2001,Martin-lopez2012} and in metrology to beat the standard shot noise limit\cite{Giovannetti2004,Buckley2012}. 

While sources of coherent light such as lasers and microwave generators are well developed, generating indistinguishable single photons on demand over a wide range of frequencies remains a challenge. Currently, prominent sources of single photons include Spontaneous Parametric Down-Conversion (SPDC) and single atomic emitters\cite{Migdall2013}. SPDC is an inherently random process that generates correlated pairs of photons (signal and idler) from a strong coherent field of higher frequency (pump) where, the detection of an idler photon heralds the signal photon. In contrast, a single emitter such as an atom, ion or quantum dot is an on-demand source which when excited by an external control relaxes by releasing the desired single photon. The problem in such a setup is the collection efficiency as the photon is radiated in all the available spatial modes. This problem has been overcome to a certain extent by embedding the emitters in a cavity and using the Purcell effect\cite{Kuhn2010}.

In the microwave regime, superconducting artificial atoms coupled to 1-D transmission line resonators have been used to efficiently generate single photons\cite{Houck2007,Hofheinz2008,Bozyigit2011}, even with controlled temporal envelopes\cite{Yin2013, Pechal2014}. However, the use of resonators in these systems reduces the bandwidth of operation. To make such a setup broadband, one has to precisely tune both the qubit and cavity frequencies which would lead to a change in the coupling strength of the qubit and the quality factor of the cavity. Thus, it is advantageous to move beyond the use of resonators for single photon generation. Previous work in this regard had suggested the use of two transmission lines, one strongly coupled and another weakly coupled to the atom, for efficient generation of single photons under certain parameter regimes\cite{Lindkvist2014,Peng2015}.

\section{Single photon generation}
\begin{figure}[]
\includegraphics[width=\columnwidth]{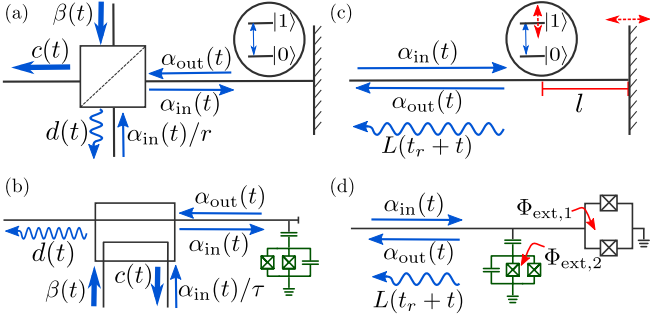}
\captionsetup{justification=justified}
\caption{Two different schematic setups for single photon generation: A coherent $\pi$-pulse $\alpha_{\text{in}}(t)$ excites a qubit at the end of a transmission line to state $\ket{1}$. The reflected field $\alpha_{\text{out}}(t)$ contains a coherent part and the emission from the qubit.  In (a), the coherent part of the reflected field is separated from the single photon using a beam-splitter with reflection coefficient $r \approx 1$. A particular implementation in superconducting circuits is shown using a directional coupler and a transmon qubit(green) in (b). In (c), the coherent part and the atomic emission are separated by tuning the coupling of the qubit to the transmission line by either changing the effective distance to the mirror or by changing the qubit frequency. Sub-figure (d) shows a corresponding implementation in superconducting circuits.}
\label{fig:setup}
\end{figure}

In this article, we propose two different setups to generate single photons using artificial atoms without resonators(Fig.\ref{fig:setup}). We solve the problem of collection efficiency by placing the atom at the end of a semi-infinite transmission line, which corresponds to an atom in front of a mirror\cite{Hoi2013,Hoi2015}. In such a configuration, all the field emitted from the atom is routed through a single output port. We will exploit this to generate a single photon by exciting the atom to the first excited state using a coherent $\pi$-pulse and then separating the atomic decay from the coherently reflected field. Thus the proposed system is all optical and does not require a separate channel such as an electrical port to excite the qubit. 

The two suggested setups differ in how we separate the emitted single photon from the coherent part. In the first setup, we use an unbalanced beam-splitter with reflection coefficient $r \approx 1$ to isolate the single photon part in one of the output modes. We believe this setup involving only a two level system and a beam-splitter is the simplest implementation of deterministic single photon generation yet. In the second setup, we tune the coupling of the qubit to the transmission line by modulating either the boundary condition or the qubit frequency. By decoupling the qubit from the transmission line at the end of the $\pi$-pulse, we isolate the single excitation in the qubit which could then be released on demand. Moreover, by tuning the coupling we could also release the photons in arbitrary wave-packets, which is essential in quantum networks for efficient information transfer.

Consider the two lowest levels of an atom (\textit{i.e.} a qubit) at a distance $l$ in front of a mirror driven by a coherent pulse $\alpha_{\text{in}}(t)$ at frequency $\omega_d$ close to the qubit frequency $\omega_{01}$. The output field using the standard input-output relation is  $\alpha_{\text{out}}(t) =  \alpha_{\text{in}}(t) e^{i\phi} + (1+e^{i\phi})\sqrt{\frac{\Gamma}{2}}\sm(t)$, where $\sm \equiv \ket{0}\bra{1}$ is the lowering operator of the qubit and $\phi = (2\omega_{01}/c) l$ is the phase gained by the field during the round-trip between the qubit and the mirror\cite{Hoi2015}. We assume an open boundary condition with the field anti-node at the end of the transmission line and define an effective coupling strength of the qubit to the transmission line, $\Gamma_{\text{eff}}(\phi) \equiv \Gamma \left(1+\cos\phi \right)$ \cite{Koshino2012}. The input that we consider here is a $\pi$-pulse $\alpha_{\text{in}}(t)=\Big[\Theta(t-t_0)-\Theta(t-t_0-t_w)\Big]\alpha_0$, where $\Theta$ is the Heaviside step function, $t_0$ is the time of arrival of the pulse to the qubit and $t_w = \pi\Big/\left(2 \alpha_0 \sqrt{\Gamma_{\text{eff}}(\phi)} \right)$ is the width of the pulse considering a constant $\phi$.

\subsection{Setup I : Using a beam-splitter}
In the first setup (Fig.\ref{fig:setup}(a) and (b)), we consider an effective coupling that is fixed by design such that $\phi$ is any value other than $\pi$. Without any loss of generality, we take $\phi=0$ and denote the effective coupling as just $\Gamma_{\text{eff}}$. The output field $\alpha_{\text{out}}$ is then interfered with a strong coherent field $\beta$ in a beam-splitter. The input-output relation for the beam-splitter with a reflection coefficient  $r$ and a $\pi/2$ phase shift in the reflected field is
\begin{align}
\left(
\begin{array}{c}
 c \\
 d \\
\end{array}
\right){=}\left(
\begin{array}{cc}
 \tau & ir \\
 ir & \tau \\
\end{array}
\right)\left(
\begin{array}{c}
\alpha_{\text{in}}(t) + \sqrt{\Gamma_{\text{eff}}} \sm(t)\\
 \beta(t)\\
\end{array}
\right),
\end{align}
where the transmission coefficient $\tau \equiv \sqrt{1-r^2}$. Choosing $\beta(t) = -i r \alpha_{\text{in}}(t) / \tau$, the coherent part of the reflected field is perfectly canceled and we have $d(t) = i r \sqrt{\Gamma_{\text{eff}}}\sm(t)$, which is our output of interest containing only the single photon. In the ideal limit with $r \to 1$ and the probability of excitation of the qubit $P_{\text{exc}} \to 1$, we will approach unit probability of having a single photon in the mode $d$, $P_1 \to 1$. Depending on the experimental implementation, one could instead chose the opposite limit with $\tau \to 1$ and isolate the single photon in the output mode $c$. While the above limits themselves are impossible to achieve, we believe that with current technologies single photon probabilities greater than $0.97$ can be easily achieved, especially using an on-chip superconducting directional coupler as beam-splitter\cite{Lang2013}. As a proof of principle, we show experimental data for cancellation of coherent pulses using 20 dB directional couplers in Appendix \ref{App:ExptResult}. We observe cancellation by a factor of 2500 (-34 dB) at room temperature.

\subsection{Setup II : Using tunable coupling}
We now move on to the second setup (Fig.\ref{fig:setup}(c) and (d)), where, the focus is to use the tunable coupling of the qubit to also shape the photon wave-packet. The operation of this scheme is as follows. We start with an initial phase $\phi_i$ close to $\pi$ either by using the external flux through the SQUID at the end of the transmission line \cite{Koshino2012} or by designing the distance to the mirror  in fabrication. We then excite the qubit using a coherent $\pi$-pulse of width $t_w$. As the effective coupling of the qubit $\Gamma_{\text{eff}}$ with this initial phase is close to zero (but not zero), the qubit doesn't relax  significantly during this pulse width. At the end of the $\pi$-pulse, we tune the external flux to have $\phi=\pi$ and decouple the qubit from the transmission line completely. We assume this can be done arbitrarily fast as we start with a phase close to $\pi$. Now the single excitation is stored in the qubit while the rest of the coherent field has been reflected back. At an arbitrary release time $t_r$, we tune the phase back  away from $\pi$ and release the excitation from the qubit. By choosing an appropriate function $\phi_r(t)$, we can change the coupling such that it also satisfies the relation $\Gamma_{\rm eff} (t) = |\xi(t)|^2 \Big/ \int_t^\infty |\xi(s)|^2 ds$, which leads to the release of the  photon in a desired wave-packet $\xi(t)$\cite{Gough2012}. We note that one could in principle achieve all of the above by tuning $\omega_{01}$ instead of having a SQUID at the end of transmission line\cite{Hoi2015}. In such a case, one could also forego the use of beam-splitter in the previous setup. Instead, by fast tuning the frequency of the qubit at the end of the $\pi$-pulse, we can separate the coherent part from the single photon with a filter since they come out at different frequencies.

\subsection{Photon generation efficiency}
We characterize both setups by calculating the probability of having $n$ photons, $P_n$ in the output field. The single photon probability $P_1$ then directly corresponds to the efficiency since an ideal single photon source would have $P_1=1$ with all other probabilities equal to 0. The photon number probabilities are calculated from the $m^{\text{th}}$ order correlation function
\begin{equation}
\label{Eq_mth_correlator}
G^{(m)}(t_1,t_2,...,t_m) \equiv \expec{L^\dagger(t_1) ...L^\dagger(t_m)L(t_m)...L(t_1)},
\end{equation}
where $L \equiv \sqrt{\Gamma_{\text{eff}}(\phi)}e^{i\phi/2}\sm$. Integrating the correlation function over all times from the release time $t_r$ to final time $T$, gives the number of photon $m$-tiples
\begin{equation}
N_m=\int_{t_r}^T dt_1 \int_{t_1}^T dt_2 ... \int_{t_{m-1}}^T dt_m G^{(m)}(t_1,t_2,...,t_m),
\end{equation}
which is related to the probabilities $P_n$ as $N_m= \sum_{n=m}^k  \binom{n}{m} P_n$ \cite{Lindkvist2014}. $k$ is the cutoff that we have chosen above which the probabilities are negligible. For the setup with the beam-splitter, we have $t_r = t_0$. 

To calculate the correlation functions we start from the master equation for the qubit  {(See Appendix \ref{App:ME_derivation} for the derivation following \cite{Hoi2015})}
\begin{equation}
\label{Eq_master_main}
\dot{\rho}(t) = -i\big[H(t),\rho(t)\big]+ \lind{L}\rho(t) \equiv \mathcal{L}(t)\rho(t).
\end{equation}
We have defined a Liouvillian $\mathcal{L}(t)$ to simplify notation. The Hamiltonian $H(t)$ in the rotating frame of the drive is \cite{Hoi2015}
\begin{equation}
H(t) = \frac{\Delta - (\Gamma/2)\sin\phi}{2} \sz  -i \left( \alpha_{\text{in}}(t) e^{i\phi} L^\dagger - h.c. \right),
\end{equation}
where $\Delta = \omega_{01}-\omega_d$ is the detuning between the qubit and the drive. The dissipation super-operator is defined as $\lind{X}P \equiv XPX^\dagger - \frac{1}{2} \left\lbrace X^\dagger X, P \right\rbrace$. 

The above master equation can be formally solved by writing a solution $\rho(t) = P(t,0) \rho(0)$, where $\rho(0)$ is the initial density matrix of the atom in front of a mirror at time $t=0$.  $P(t,0)$ is the propagator that evolves the qubit state from  time $0$ to $t$. Substituting the above solution in the master equation \eqref{Eq_master_main}, we get $\dot{P}(t,0)=\mathcal{L}(t) P(t,0)$, with the initial condition, $P(0,0)=1$. Considering the two level system to be driven by a coherent pulse between $t_0$ and $t_w$, the solution of this equation is simply 
\begin{widetext}
\begin{align}
P(T,0) &= \exp\left(\int_{t_w}^T \mathcal{L}_0 d\tau'' \right)  \exp\left(\int_{t_0}^{t_w} \mathcal{L}_\alpha(\tau) d\tau' \right)\exp\left(\int_{0}^{t_0} \mathcal{L}_0 d\tau \right) \\
&= \exp\bigg( \mathcal{L}_0 (T-t_w) \bigg)  \exp\left(\int_{t_0}^{t_w} \mathcal{L}_\alpha(\tau)  d\tau' \right)\exp\left( \mathcal{L}_0 t_0 \right),
\end{align}
\end{widetext}
where $\mathcal{L}_0 = \mathcal{L}(\alpha_{\text{in}}=0)$ is the Liouvillian without any driving and $\mathcal{L}_\alpha$ contains the time-dependent driving term. Considering a square pulse that has a constant value $\alpha_0$ between $t_0$ and $t_w$ and 0 everywhere else, the integral containing the drive can also be evaluated immediately. Any two time propagator $P(t_2,t_1)$ can be evaluated similarly, taking into account in which regions $t_1$ and $t_2$ fall. The $m^{\text{th}}$ order correlation function defined in \eqref{Eq_mth_correlator} can then be calculated using the propagator as \cite{Gardiner2004}
\begin{widetext}
\be
G^{(m)}(t_1,t_2,...,t_m) = \Tr \Bigg[L P(t_{m-1},t_m)...\Bigg\lbrace L P(t_3,t_2)\bigg\lbrace L P(t_2,t_1)\Big\lbrace L\rho(t_1)L^\dagger \Big\rbrace L^\dagger \bigg\rbrace L^\dagger \Bigg\rbrace...L^\dagger \Bigg].
\ee
\end{widetext}
From the correlation functions, we can calculate the photon m-tiples and the photon probability distribution as explained above.

\begin{figure}
\includegraphics[width=\columnwidth]{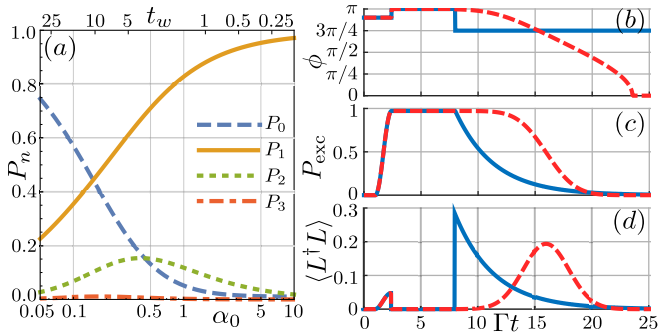}
\captionsetup{justification=justified}
\caption{\textbf{(i)} Photon generation using a beam-splitter: (a) Photon number distribution in the output field after the beam-splitter as a function of the input drive strength $\alpha_0$ of the $\pi$-pulse with $r=0.995$. The incoming field is on resonance with the qubit \textit{i.e.} $\Delta=0$. All the parameters are in the units where $\Gamma_\text{eff}=1$. For $\alpha_{0}=5$ we have a $P_1 \approx 0.95$ and for $\alpha_{0}=10$ we have a $P_1 \approx 0.97$. \textbf{(ii)} Generating a single photon with arbitrary wave-packet envelopes using tunable coupling: (b) $\phi$ at different time steps, (c) corresponding probability of excitation $P_{exc}$ of the qubit and (d) the average flux output from the qubit $\expec{L^\dagger L}$. We have shown the results for two different wave-packets: an exponential (blue, solid) and a Gaussian (red, dashed). We have the initial phase $\phi_i=0.9 \pi$ and $\phi_r(t)$ is chosen to be get the corresponding wave-packets. The $\pi$-pulse starts at $t_0=1$ with $\alpha_0=5$ and the photon is released from $t_r=8$. The scale is set by $\Gamma=1$.  From this simulation, we find that between $t_r$ and $T=20$, we have $P_1 \approx 0.97$. For $\alpha_0 = 10$, we correspondingly find $P_1 \approx 0.99$.}
\label{fig:PhotonProbabilities}
\end{figure}

\subsection{Results}
Using the above expressions, we can now evaluate the efficiency of both setups, starting  with the one containing the beam-splitter. The photon probabilities in the output field from the beam-splitter is plotted in  Fig.\ref{fig:PhotonProbabilities}(a) as a function of the input amplitude, where we have taken a cut-off of $k=3$. At very low input powers close to 0, the atom is rarely excited leading to only vacuum in the output field. At higher input powers, we have significant excitation of the qubit and $P_0$ falls towards 0. However if the $\pi$-pulse is long, the qubit can be excited again after an initial decay during the pulse width. This leads to a non-negligible two-photon probability $P_2$. As $\alpha_0$ increases and $t_w$ decreases, the higher photon probabilities vanishes and we are left with an efficient single photon source. In these calculations, we have used $r = 0.995$ to get a single photon with $\sim$95\% efficiency using $\alpha_0=5$ and with $\sim$97\% efficiency using $\alpha_{0}=10$. We believe these values are straightforward to achieve with current technologies. 

Next, we turn to the setup with tunable coupling and focus on generating single photon wave-packets with an envelope that can be shaped in time. In Fig.\ref{fig:PhotonProbabilities}, we show the results for two different wave-packets: an exponential and a Gaussian. The exponential wave-packet is naturally generated using a constant coupling with any $\phi_r \neq \pi$. To get the Gaussian wave-packet, the coupling has to be tuned in a non-trivial fashion as shown by the red dashed line in Fig.\ref{fig:PhotonProbabilities}(b). The width of the wave-packets are limited by the maximum ($2\Gamma$) and minimum (0) coupling that we can reach. The efficiency of the single photon generation in this case is limited by the loss of excitation during the $\pi$-pulse, which can be seen as a small bump in the photon flux before $t_r$. We see that with an initial  phase $\phi_i=0.9 \pi$, we can reach single photon  generation efficiency of $\sim97\%$ using $\alpha_0=5$ and $\sim 99\%$ using $\alpha_0=10$.

\subsection{Limitations}
We now look at some of the limitations coming from the qubit on the presented schemes. The width of the $\pi$-pulse in  both setups is limited by the anharmonicity $\omega_{01}-\omega_{12}$ of the qubit. To confine ourselves to the lowest two levels of the atom, we require that the Rabi frequency of the drive $\Omega = 2\alpha_0 \sqrt{\Gamma_{\text{eff}}}$ be less than the anharmonicity. For a transmon with typical anharmonicity of about 250-300 MHz and effective coupling strength of about 60-80 MHz\cite{Hoi2013}, we require $\alpha_0$ to be less than 20-30 which is  higher than the values we have considered here. By using qubits with higher anharmonicity such as flux qubits\cite{Mooij1999}, one could push the drive strength much further, taking the single photon probability even closer to 1.  Next we look at the effect of pure dephasing of the qubit, which we have neglected so far in the discussion. In the setup with the beam-splitter, the efficiency of single photon generation is impacted by dephasing only during the $\pi$-pulse, which can be made short by increasing $\alpha_0$ subject to the limit from anharmonicity. In the setup with the tunable coupling, if dephasing leads to a change in energy splitting of the qubit, it would also  lead to a slow variation of $\phi$. Then, the excitation stored in the qubit has to be released more quickly than the scale set by the dephasing time to avoid additional losses. In engineered quantum systems such as superconducting circuits, the dephasing rates have been made much smaller compared to the coupling strengths over the past few years and the total decoherence is often $T_1 (\propto 1/\Gamma)$ limited \cite{Devoret2013}. Thus, we believe the proposed setups are not significantly affected by dephasing. We also believe that with two independent photon sources with low dephasing rates, we can generate indistinguishable photons and also create entangled states using the Hong-Ou-Mandel effect with high efficiency similar to \cite{Lang2013}. In all of the above calculations, we have  neglected other decay channels usually referred to as non-radiative decay channels. The effect of these other channels can be made negligibly small for superconducting circuits as shown by the extinction of forward scattering of coherent light in \cite{Astafiev2010a,Hoi2012} and hence are not important for our current proposal. However, to keep the discussion complete we analyze the effect of other channels on the photon generation efficiency in Appendix \ref{App:NRdecay}.

\subsection{Flying qubits}
Photons are prime candidates to act as flying qubits that transfer information across quantum networks, as they rarely interact with each other. The proposed setups above can also be used as sources for generating flying qubits of the form $\mu\ket{0} + \nu\ket{1}$  as follows. Any arbitrary state of the above form can be encoded on the qubit using the detuning $\Delta$, amplitude and phase of the drive $\alpha_{in}$ and pulse width of the drive $t_w$. Via the coupling to the transmission line, the qubit state can be transferred to the field  and then be separated from the coherent part as before. With the tunable coupling, the state can also be encapsulated inside a wave-packet that can then be efficiently captured by the receiver\cite{Yin2013}. 

\section{Generation of correlated photons}

\begin{figure}[]
\includegraphics[width=\columnwidth]{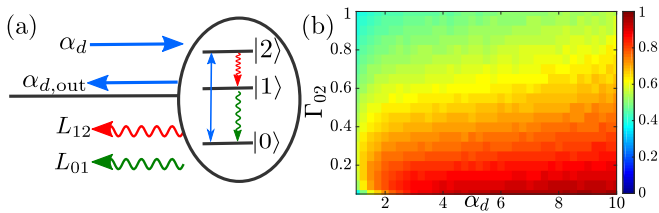}
\captionsetup{justification=justified}
\caption{(a) Schematic setup for cascaded pair photon production and (b) measure of correlation between the fields from $0-1$ and $1-2$ transition, $V$ as a function of $\alpha_d$ and $\Gamma_{02}$. We work in units where $\Gamma_{01}=1$ and with $\Gamma_{12}=2\Gamma_{01}$. The drive $\alpha_d$ is on resonance with the $0-2$ transition and the atom is placed at the end of the transmission line i.e. with $\phi=0$. $V > 0$ implies non-classical correlations. We find that with $\Gamma_{02}/\Gamma_{01} \approx 1/20$ (reasonable value for a transmon) and with reasonable drive strength $\alpha_d=5$, we have $V \approx 0.92$, very close to the maximum value of 1.}
\label{fig:Correlated_setup}
\end{figure}

Given the simplicity of the setups discussed above, we can immediately extend them to include higher levels of the atom, to generate correlated photons from a cascade process (Fig. \ref{fig:Correlated_setup}). We once again consider an atom in front of a mirror with $\phi=0$ for simplicity. A coherent $\pi$-pulse $\alpha_d(t)$ resonant with the $0-2$ transition, excites the atom to the state $\ket{2}$. If the coupling strength of the $0-2$ transition is weak $\Gamma_{02} \ll \Gamma_{12},\Gamma_{01}$, the atom will preferably relax by the cascade process. In this setup, we do not employ any beam-splitters as the coherent pulse and the desired correlated photons are at different frequencies and can be easily separated by filtering.

To quantify the correlations between the pair of photons, we define the function $V \equiv G_{is}^2-G_{ii}G_{ss}$, where 
\begin{equation}
G_{ab} \equiv \int_{0}^{T} \int_{0}^T dt_1 dt_2 \expec{L_a^\dagger(t_1) L_b^\dagger(t_2) L_b(t_2) L_a(t_1)},
\end{equation}
with $L_{01/12} = \sqrt{\Gamma_{01/12}}\sm^{(01/12)}$. $V$ is negative for classical beams and becomes positive for two beams with quantum correlations, reaching a maximum value of 1 for a perfect pair production. The above quantity is also related to the Cauchy-Schwarz inequality (CSI) \cite{Clauser1974,Walls2008} defined here as $V<0$, which is violated for two beams with quantum correlations. The value of $V$ is shown in Fig.\ref{fig:Correlated_setup}(b) as a function of the drive strength and the $0-2$ coupling strength $\Gamma_{02}$. As one would expect, the correlations are much stronger for small $\Gamma_{02}$ and larger $\alpha_d$. In these simulations, we have considered the transmon limit with $\Gamma_{12}=2\Gamma_{01}$. For a transmon qubit, $\Gamma_{02}$ is expected to go to 0 as $E_J/E_C \to \infty$ \cite{Koch2007}. However, in practical devices, we have $\Gamma_{02}/\Gamma_{01} \approx 1/20$\cite{Peterer2015}. With this coupling strength and with reasonable drive strength $\alpha_d=5$, we have $V \approx 0.92$.  Such correlated photon pair sources are useful as heralded single photon sources \cite{ZelDovich1969,Hong1986} and for calibrating photon detectors\cite{ZelDovich1969,Polyakov07}. While other sources of correlated microwave photons have already been  demonstrated \cite{Wilson2011,Flurin2012}, we emphasize that the setup discussed here is non-stochastic.

\section{Summary}
In summary, we have proposed simple and efficient setups to generate microwave photons on chip. Such setups along with others based on voltage biased Josephson junctions\cite{Leppakangas2015} offer photon generation protocols not limited by resonator bandwidths. By tuning the frequency of the qubits in-situ, photons over a wide range of frequencies can be generated in the proposed setups. While we have focused on microwave photons and superconducting circuits in this article, we believe it is straightforward to implement these proposals in other solid-state devices such as quantum dots coupled to one dimensional waveguides\cite{Arcari2014}, atomic emitters coupled to silica nanofibers\cite{Chonan2014} for generating either optical or microwave photons.

\begin{acknowledgments}
We thank Joel Lindkvist for valuable discussions. We acknowledge financial support from the Wallenberg Foundation and  the European Union through the EU STREP project PROMISCE.
\end{acknowledgments}

\appendix
\section{Master equation of an atom in front of a mirror}
\label{App:ME_derivation}
In this section, we briefly review the derivation of the master equation for an atom in front of a mirror following \cite{Hoi2015}. To do so, we use a convenient formalism known as the $(S,L,H)$ formalism \cite{GoughCommMathPhys2009,GoughIEEE2009}. In this formalism, a quantum system is described by a triplet
\be
G\equiv(S,L,H),
\ee
where $S$ is the scattering matrix, $L$ is the vector of  coupling operators and $H$ is the Hamiltonian. We then define the following three operations for composing multiple quantum systems into one system. 
\begin{widetext}
\begin{enumerate}
\item The series product $\triangleleft$ of the triplets is used to denote the feeding of output of one subsystem to another
\begin{eqnarray}
G_2 \triangleleft G_1 = \Bigg(S_2 S_1, S_2L_1 + L_2, H_1 + H_2 + \frac{1}{2i}\left(L_2^\dag S_2 L_1 - L_1^\dag S_2^\dag L_2 \right) \Bigg).
\label{seriesproduct} 
\end{eqnarray}
\item The concatenation product $\boxplus$ is used for composing subsystems into a system with	 stacked channels
\begin{eqnarray}
G_2 \boxplus G_1 &=& \left( \begin{pmatrix} S_2 & 0 \\ 0 & S_1 \end{pmatrix}, \begin{pmatrix} L_2 \\  L_1 \end{pmatrix}, H_2 + H_1 \right).  
\label{catproduct} 
\end{eqnarray}
\item Finally, we also have a operation for feedback, written as $[(S,L,H)]_{k\rightarrow l} = (\tilde{S},\tilde{L},\tilde{H}),$ where the output from the $k^{th}$ port of the system is fed back as the input through the $l^{th}$ port of the same system. The triplet is given by
\begin{align}
\label{Eq:SLH_feedback_rules}
\tilde{S} &= S_{[\cancel{k,l}]} + \begin{pmatrix}
S_{1,l} \\
\vdots \\
S_{k-1,l} \\
S_{k+1,l} \\
\vdots \\
S_{n,l}
\end{pmatrix}
\left(1-S_{k,l}\right)^{-1}
\begin{pmatrix}
S_{k,1}\:
\dots \:
S_{k,l-1} \:
S_{k,l+1} \:
\dots \:
S_{k,n}
\end{pmatrix},
\nonumber\\
\tilde{L} &= L_{[\cancel{k}]} + 
\begin{pmatrix}
S_{1,l} \\
\vdots \\
S_{k-1,l} \\
S_{k+1,l} \\
\vdots \\
S_{n,l}
\end{pmatrix}
\left(1-S_{k,l}\right)^{-1}L_k,
\nonumber\\
\tilde{H} &= H + \frac{1}{2i}\left(\left(\sum_{j=1}^n L_j^\dag S_{j,l} \right)\left(1-S_{k,l}\right)^{-1}L_k - \text{h.c.} \right),
\end{align}
where $S_{[\cancel{k,l}]}$ and $L_{[\cancel{k}]}$ are the original scattering matrix and coupling vector with row $k$ and column $l$ removed \cite{Tezak2012}.
\end{enumerate}
\end{widetext}
Using the above defined operations, we can write down the $(S,L,H)$ triplet for the whole system 
\be
G_{\rm tot} = \left(S_{\rm tot}, \begin{pmatrix} L_1 \\ \vdots \\ L_n \end{pmatrix}, H_{\rm tot}\right).
\ee
The master equation is written from the above total triplet as 
\be
\label{Eq:MEfromSLH}
\dot{\rho} = -i\comm{H_{\rm tot}}{\rho} + \sum_{i=1}^n \lind{L_i}\rho,
\ee
where the dissipation super-operator is given by $\lind{X} \rho = X\rho X^{\dagger }-\frac{1}{2}X^{\dagger }X\rho -\frac{1}{2} \rho X^{\dagger }X$. The output from the $i$th channel is simply given by $L_i$.

\begin{figure}
\includegraphics[width=\columnwidth]{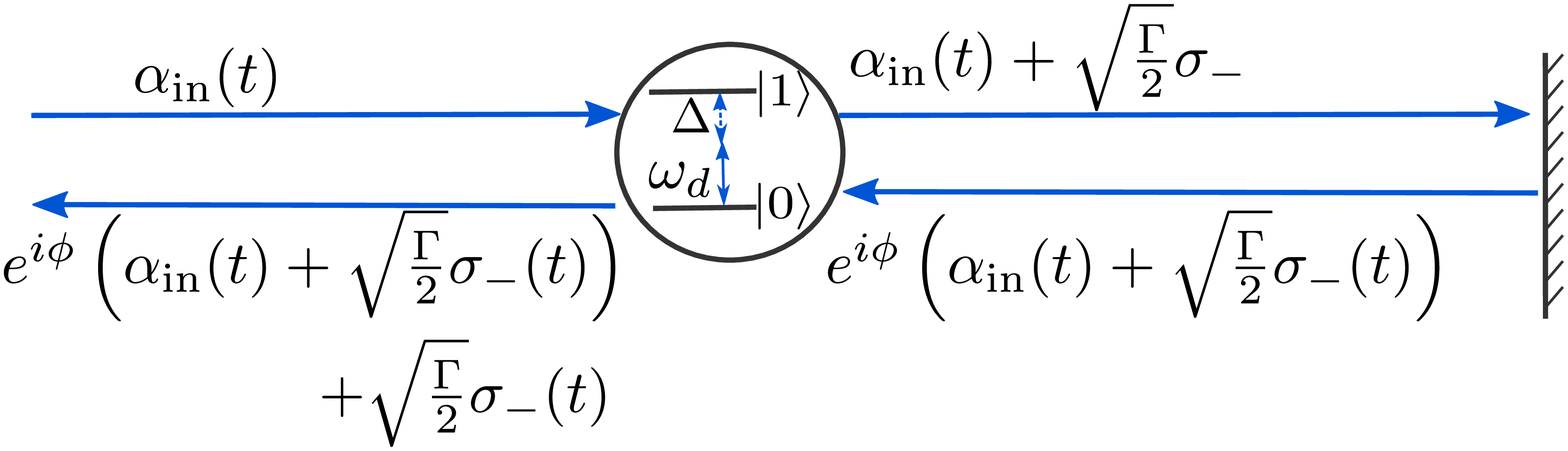}
\caption{A two level system in front of a mirror. The two level system is driven from the left, by a coherent field $\alpha_{\text{in}}(t)$ of frequency $\omega_d$. The distance between the qubit and the mirror is modeled as a phase $\phi$ gained by the field during the round-trip.}
\label{fig:detailed}
\end{figure}

We can now use the above rules and derive the master for a two level system that is in front of a mirror. A schematic is shown in Fig.\ref{fig:detailed}, where we have also marked all the inputs and outputs. The distance between the qubit and the mirror is modeled as a phase $\phi$  gained by the field. The field reflected from the mirror is fed back to the qubit and we can use the feedback operation of the $(S,L,H)$ formalism to derive the master equation as in \cite{Hoi2015}. To keep the notations simple, we drop the time-dependence of the operators in the following. 

The triplet for a two-level atom coupled to an open transmission line (i.e. with both left and right propagating modes) without any drive is given as
\be 
G_{\text{TLS}} = \left(\mathbb{1}, \begin{pmatrix} \sqrt{\frac{\Gamma}{2}} \sm \\ \sqrt{\frac{\Gamma}{2}} \sm \end{pmatrix}, H_{\text{TLS}} \right). 
\ee
The Hamiltonian of the two level system $H_{\text{TLS}} = -(\omega_{01}/2) \sz$, where $\omega_{01}$ is the energy difference between the $\ket{0}$ and $\ket{1}$ state of the qubit. $\Gamma$ is the coupling strength of the qubit to the transmission line and  $\sm=\sp^\dagger=\ket{0}\bra{1}$ is the lowering operator of the qubit. The triplet for the phase shift gained by the field is
\be 
G_{\phi} = (e^{i\phi},0,0).
\ee 
 With these, we can write the total triplet for the qubit-mirror without any drive as
\be
G = (\tilde{S},\tilde{L},\tilde{H})= [(G_{\phi} \boxplus I) \triangleleft G_{\text{TLS}}]_{1\rightarrow 2}.
\ee 
We have to concatenate an identity triple $I=(1,0,0)$ to keep the same dimensions. Using the feedback rules given in \eqref{Eq:SLH_feedback_rules}, we get
\begin{align} 
\tilde{S} &= e^{i \phi}, \\
\tilde{L} &= \sqrt{\frac{\Gamma}{2}}(1 + e^{i \phi}) \sm = e^{i\phi/2} \sqrt{\Gamma_{\text{eff}}} \; \sm, \\
\tilde{H} &= H_{\text{TLS}} + \frac{\Gamma}{2} \sin\phi\; \sp\sm,
\end{align}
where $\Gamma_{\text{eff}} \equiv \Gamma (1+\cos\phi)$. With this triplet, we can now include the coherent drive in its own rotating frame specified by a triplet 
\be 
G_\alpha = (1,\alpha_{\text{in}},0).
\ee
The total triplet is then
\be
G_{\rm tot} = (\tilde{S},\tilde{L},\tilde{H}) \triangleleft G_{\alpha}
\ee
which gives the master equation
\be
\dot{\rho}= -i[H_{\rm tot},\rho] + \lind{e^{i\phi}\alpha_{\text{in}} + e^{i\phi/2} \sqrt{\Gamma_{\text{eff}}} \; \sm}\rho, 
\ee
where the Hamiltonian 
\be
H_{\rm tot} = \frac{\Delta}{2} \sz + \frac{\Gamma}{2} \sin\phi\; \sp\sm + \frac{1}{2i} \left(e^{i\phi/2} \sqrt{\Gamma_{\text{eff}}} \; \alpha_{\text{in}}   \sp - h.c. \right).
\ee
The Hamiltonian is also in the rotating frame of the incoming field with $\Delta = \omega_{01}-\omega_d$. We can expand the dissipator and simplify the master equation to get
\be
\dot{\rho}= -i[H_{\rm eff},\rho] + \lind{L}\rho \equiv \mathcal{L} \rho, 
\label{Eq_master}
\ee
where $L=\sqrt{\Gamma_{\text{eff}}}e^{i\phi/2}\sm$ and the effective Hamiltonian is
\be
H_{\rm eff} = \frac{\Delta}{2} \sz +\frac{\Gamma}{2} \sin\phi\; \sp\sm   - i \left(e^{i\phi} \alpha_{\text{in}} L^\dagger - h.c. \right).
\ee
To keep the notations simple, we have also defined a Liouvillian $\mathcal{L}$. By rewriting $\sp\sm=  (\mathbb{1}-\sz)/2$ and neglecting the constant term, we end up with the Hamiltonian given in Eq.(5) of the main text.

\section{Effect of other decay channels}
\label{App:NRdecay}
In all of the results presented in the main text, we have not considered the effect of channels other than the transmission line on the efficiency of single photon generation. As mentioned there, the systems we considered are with near unity coupling to the transmission line where the effect of these other channels is negligibly small. However, to keep the discussion complete, here we calculate the efficiency of single photon generation in systems where the effect of these other (non-radiative) decay channels are not negligible. To do so we modify the master equation \eqref{Eq_master} as 
\be
\dot{\rho}= -i[H_{\rm eff},\rho] + \lind{L}\rho + \Gamma_{\text{nr}} \lind{\sm} \rho, 
\ee
where $\Gamma_{\text{nr}}$ is the effective coupling strength of the qubit to all of the decay channels other than the transmission line. 

\begin{figure}
\captionsetup{justification=centerlast, singlelinecheck=false}
  \includegraphics[width=\linewidth]{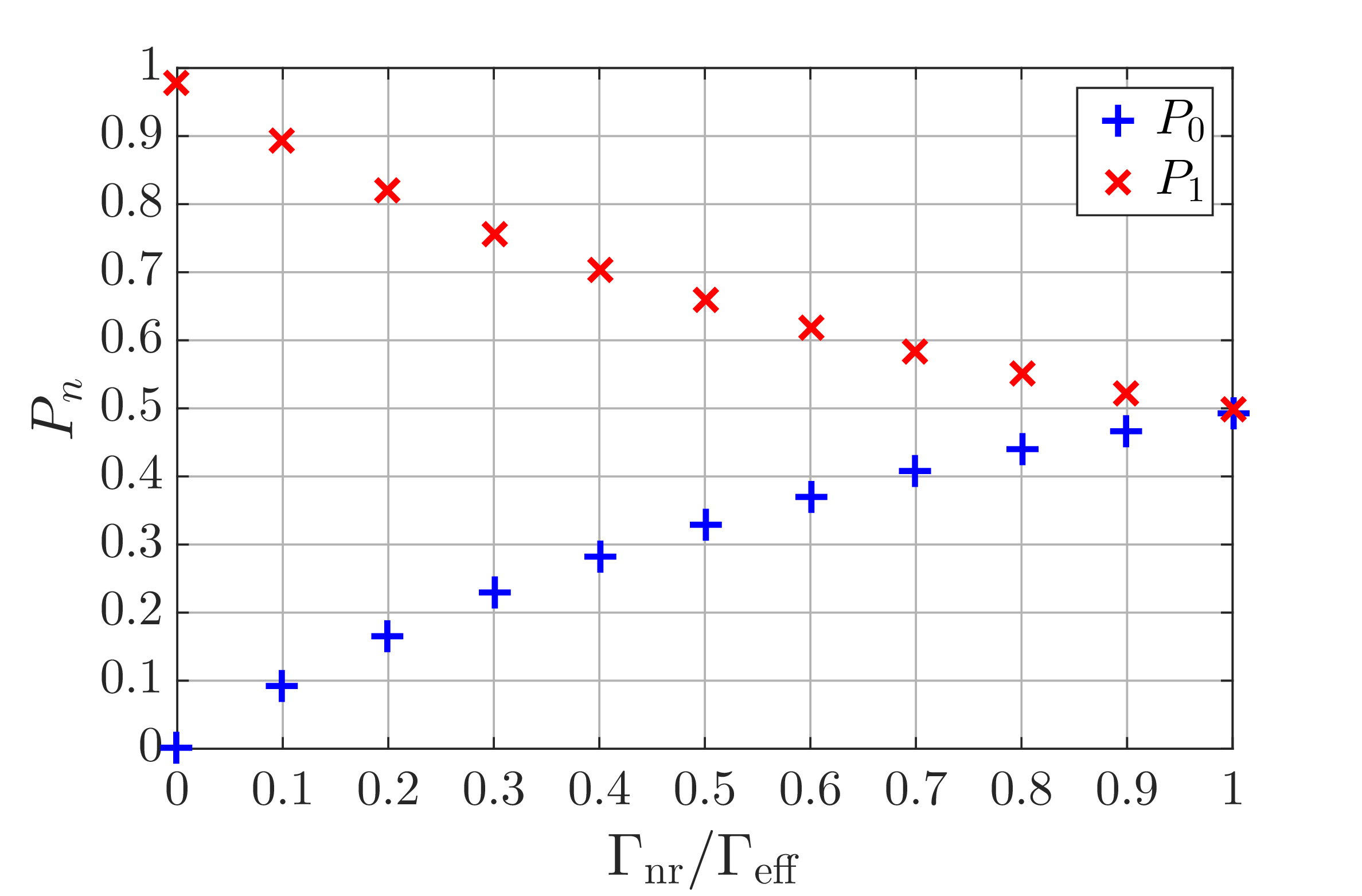}
  \caption{Photon generation using a beam-splitter: Probability of having 0 and 1 photons in the output field in the presence of additional decay channels. The effect of the other decay  channels is given by their total coupling strength to the qubit $\Gamma_{\text{nr}}$. Here we have considered the input drive strength $\alpha_0 = 10$ and the reflection coefficient of the beam-splitter $r=0.995$. The incoming field is on resonance with the qubit \textit{i.e.} $\Delta=0$. All the parameters are in the units where $\Gamma_\text{eff}=1$. As expected, the efficiency of single photon generation goes down with increasing $\Gamma_{\text{nr}}$.}
  \label{fig:PvsGammaNR}
  \end{figure}

For the setup with the beam-splitter, we calculate the output photon probabilities as outlined in the previous section by solving the above master equation. We show the probability to have 0 and 1 photons in the output in Fig.\ref{fig:PvsGammaNR}. As one would expect, the efficiency of single photon generation goes down with additional losses coming from these other channels. When $\Gamma_{\text{nr}} = \Gamma_{\text{eff}}$, we get the efficiency of around $50\%$ that one would expect from having a qubit in an open transmission line \cite{Lindkvist2014}. 

\begin{figure}
  \includegraphics[width=1\linewidth]{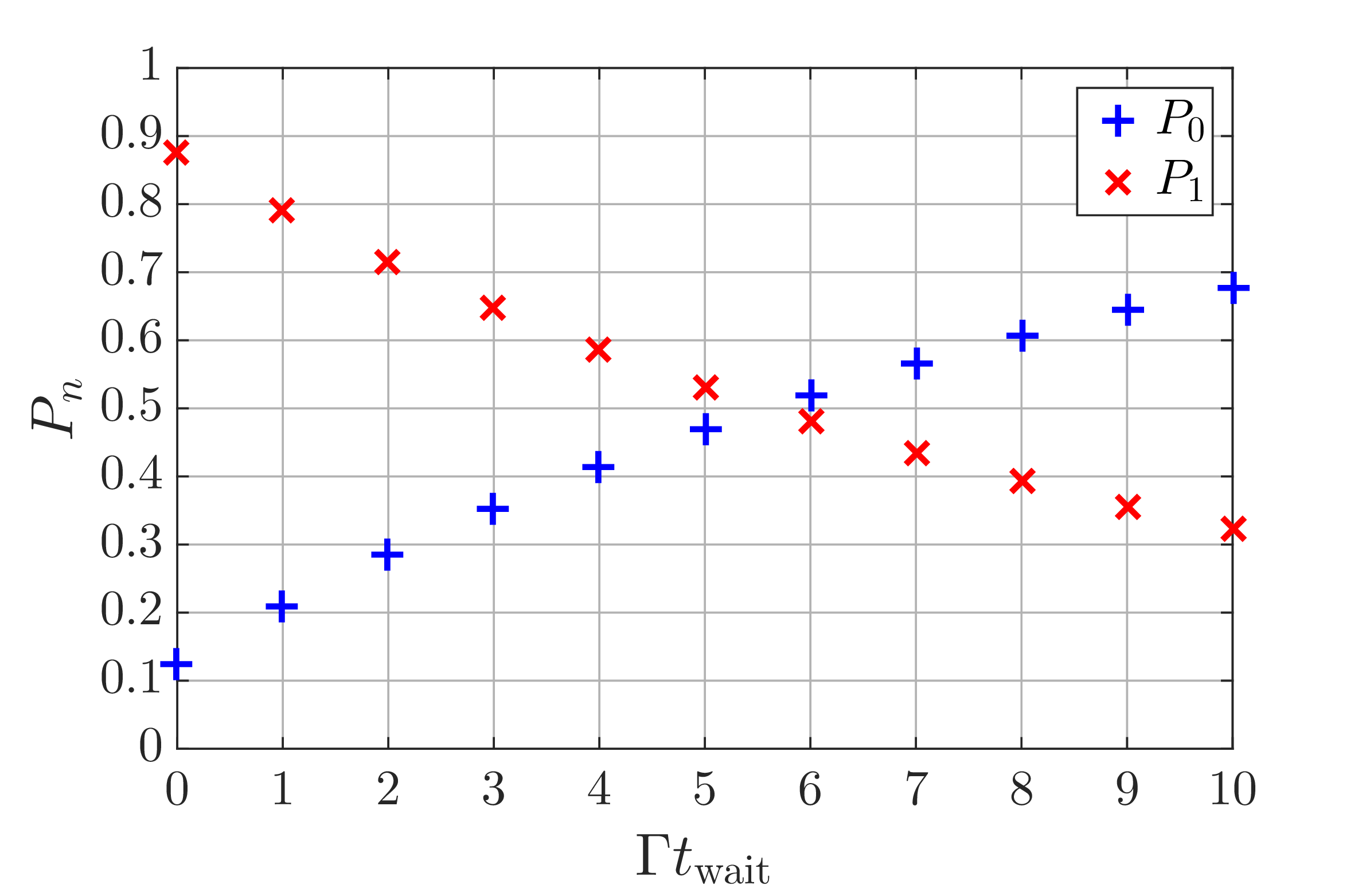}
  \captionsetup{justification=centerlast, singlelinecheck=false}
  \caption{Photon generation using a tunable coupling: Probability of having 0 and 1 photons in the output field in the presence of additional decay channels as a function of delay time between the $\pi$-pulse and the relaxation of the qubit. $t_{\text{wait}}=t_r - (t_0+t_w)$. Here we consider the coupling of the qubit to other decay channels $\Gamma_{\text{nr}} = 0.1 \Gamma$. The input drive strength $\alpha_0 = 10$ and phase during the release $\phi_{r}=\pi/2$. The incoming field is on resonance with the qubit \textit{i.e.} $\Delta=0$. All the parameters are in the units where $\Gamma=1$. As expected, the efficiency of single photon generation goes down as we wait longer.}
  \label{fig:Pvstwait}
\end{figure}

For the setup using the tunable coupling, the presence of the other channels can lead to reduction of photon generation efficiency due to loss of excitation during the storage time or during the relaxation. Thus, the excitation trapped in the qubit has to be released faster and also in shorter pulse widths to reduce significant losses. In Fig.\ref{fig:Pvstwait}, we show how the efficiency of photon generation scales with the waiting time i.e. the time between end of $\pi$-pulse and the release, $t_{\text{wait}}=t_r - (t_0+t_w)$, for a particular $\Gamma_{\text{nr}}$. As expected the more we wait, there is more chance of losing the excitation in other channels.

From the above analysis we see that, even with $\Gamma_{\text{nr}}$ around 10 to 20\% of $\Gamma$, single photon generation efficiencies of  80 to 90\% are easily achievable.


\section{Measured cancellation of microwave signals using a directional coupler}
\label{App:ExptResult}
Assume two incident signals on a directional coupler, $\alpha = A_1 e^{i(\omega_1 t + \phi_1)}$ and $\beta = A_2 e^{i(\omega_2 t+\phi_2)}$, where $\omega_{1/2}$ are the angular frequencies, $A_{1/2}$ are the amplitudes and $\phi_{1/2}$ are the phases of the two signals. By putting a mirror at the third port, we can write the outgoing signal at the fourth port as
\begin{equation}
d = \tau_1 e^{i\phi} \alpha + \tau_2 \beta = \tau_1 A_1 e^{i(\omega_1 t+\phi_1+\phi)}+\tau_2A_2 e^{i(\omega_2 t +\phi_2)},
\end{equation}
where $\tau_{1/2}$ are the transmission amplitudes, which for an ideal beam splitter should be the same in the two arms. $\phi$ is the phase acquired by the signal $a$ as it travels to and from the mirror. Assuming $\tau_1=\tau_2=\tau$, perfect cancellation occurs at $d=0$ yielding 
\begin{align}
A_2&=A_1,\nonumber\\
 \phi_2&=\phi_1+\phi+(2n-1)\pi,\nonumber\\
\omega_2&=\omega_1,
\end{align}
where $n$ is an integer. The upper limit of cancellation is set by how well we control the amplitude, phase and frequency of the two signals. For pulses, the timing also has to be controlled.

\begin{figure}
    \centering
       \includegraphics[width = 0.6\linewidth]{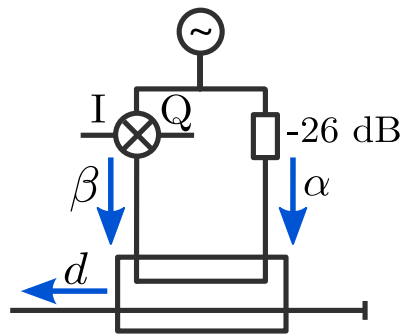}
    \caption{Schematic of the experimental setup used to demonstrate the cancellation of two microwave signals using a directional coupler. The amplitude and phase of $\beta$ is adjusted using the in-phase, I, and quadrature, Q, voltages. The right arm is attenuated by 26 dB to put the required amplitude of $\beta$ in the linear regime of the IQ-mixer.}
    \label{setup}
\end{figure}

\begin{figure*}
    \centering
       \includegraphics[width = 0.9\linewidth]{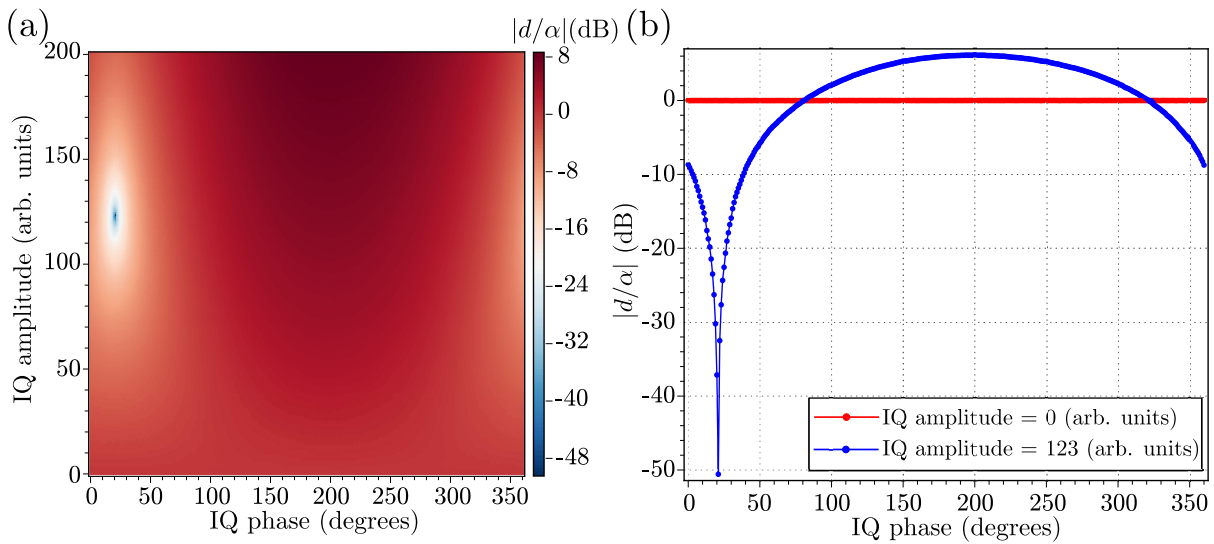}
    \caption{(a) Amplitude of output signal $d$ normalized to the amplitude of $\alpha$ as a function of phase and amplitude of the IQ-mixer. The frequency of the signal is $\omega/2\pi=5$ GHz. (b) Line cuts at two different IQ amplitudes, corresponding to the IQ mixer being on (blue) and off (red). A cancellation of maximum -50 dB is reached.}
    \label{cancel}
\end{figure*}

To avoid using two microwave sources with the possibility of frequency drift between the two signals, we use one source which is then split into two arms by a 50/50 power splitter. Arm number one, $\alpha$, is left unchanged; while arm number two, $\beta$, goes into an IQ-mixer, enabling control of both the amplitude and phase of the incoming signal using two DC-voltages corresponding to the in-phase and quadrature components,  see Fig.\ref{setup}. Then we interfere the two signals on a 20 dB directional coupler ($\tau=0.1$), and measure the outgoing amplitude and phase using heterodyne detection. The outcome of the measurement can be seen in Fig.\ref{cancel}. A maximum cancellation of -50 dB was reached at a frequency of $\omega/2\pi=5$ GHz, a typical frequency for superconducting qubits.

\begin{figure*}
    \centering
       \includegraphics[width = 0.8\linewidth]{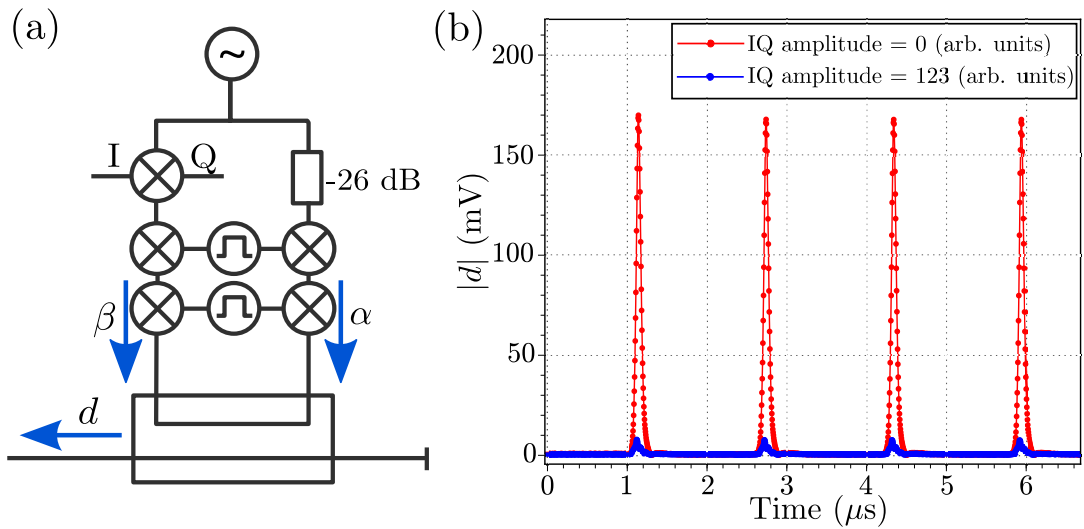}
    \caption{(a) The setup used for pulsed measurements. Both arms have two microwave mixers to shape the outgoing signal. By sharing the same AWG channels between the two arms we ensure close to identical pulses in both arms. (b) Cancellation of 100 ns Gaussian pulses. In the red trace the signal through the IQ mixer is pinched off and we only measure the pulse. In the blue trace the IQ mixer is tuned into the cancellation point and we observe up to -34 dB cancellation in the pulse.}
    \label{pulse}
\end{figure*}

To create the pulses needed to drive the qubit we install four  mixers, two in each arm. They are connected to an arbitrary waveform generator with a sampling rate of 1.25 GHz (see Fig.\ref{pulse} (a)). We use two mixers instead of one in each arm to get a better on/off ratio. To demonstrate the cancellation of microwave pulses we create four Gaussian pulses with a width of 100 ns each. To measure the cancellation we compare the outgoing field amplitude $|d|$ when the IQ mixer is turned on and off. The result is shown in Fig.\ref{pulse} (b). We observe up to -34 dB of cancellation inside the pulse.

\bibliography{SPG.bib}

\end{document}